\documentclass[a4paper,12pt,reqno]{article}
\usepackage{amsmath,amsfonts,amssymb,amsthm,dsfont}

\usepackage{hyperref}
\allowdisplaybreaks

\newcommand{\om}{\omega}

\newcommand{\6}{\partial}

\newcommand{\ua}{{\underline{\alpha}}}
\newcommand{\ub}{{\underline{\beta}}}

\newcommand{\ul}[1]{{\underline{#1}}}

\newcommand{\com}[2]{[\,#1\, ,\,#2\,]}	

\newcommand{\gam}{\Gamma}	
\newcommand{\CC}{C}	
\newcommand{\IC}{C^{-1}}	
\newcommand{\unit}{\mathds{1}}

\newcommand{\Dim}{D} 
\newcommand{\STACK}[2]{\stackrel{#1}{#2}}

\newcommand{\Then}{\Rightarrow}

\renewcommand{\mu}{m}
\renewcommand{\nu}{n}
\renewcommand{\rho}{r}
\renewcommand{\sigma}{s}
\renewcommand{\kappa}{k}
\renewcommand{\lambda}{\ell}

\begin{document}

\begin{flushright}
ITP-UH-24/16
\end{flushright}

\begin{center}
 {\large\bfseries Counterpart of the Weyl tensor for Rarita-Schwinger type fields}
 \\[5mm]
 Friedemann Brandt \\[2mm]
 \textit{Institut f\"ur Theoretische Physik, Leibniz Universit\"at Hannover, Appelstra\ss e 2, 30167 Hannover, Germany}
\end{center}

\begin{abstract}
In dimensions larger than 3 a modified field strength for Rarita-Schwinger type fields is constructed whose components are not constrained by the field equations. In supergravity theories the result provides a
modified (supercovariant) gravitino field strength related by supersymmetry to the (supercovariantized) Weyl tensor. 
In various cases, such as for free Rarita-Schwinger type gauge fields and for gravitino fields in several supergravity theories, the modified field strength coincides on-shell with the 
usual field strength.
A corresponding result for first order derivatives of Dirac type spinor fields is also presented.
\end{abstract}


This note relates to Rarita-Schwinger type fields in $\Dim$ dimensions with $\Dim> 3$. The components of these fields are denoted by $\psi_\mu{}^\ua$ where $\mu=1,\dots,\Dim$ is a vector index and $\ua=\ul 1,\dots,\ul{2^{\lfloor \Dim/2 \rfloor}}$ is a spinor index, ${\lfloor \Dim/2 \rfloor}$ being the largest integer $\leq\Dim/2$. The fields are subject to field equations (equations of motion) which read or imply $\6_\mu\psi_\nu\gam^{\mu\nu\rho}+\ldots\approx 0$ where, both here and below, ellipses denote terms that may or may not be present (but do not contain derivatives of $\psi_\mu$), $\approx$ denotes equality on-shell (i.e. equality whenever the field equations hold), spinor indices have been suppressed (as will be often done below) and $\gam^{\mu\nu\rho}=\gam^{[\mu}\gam^\nu\gam^{\rho]}$ is the totally antisymmetrized product of three gamma-matrices.\footnote{We use conventions as in \cite{Brandt:2009xv}.}
The field equations thus constrain the usual field strength whose components are
 \begin{align}
  T_{\mu\nu}{}^\ua=\6_\mu\psi_\nu{}^\ua-\6_\nu\psi_\mu{}^\ua+\ldots
	\label{eq2} 
 \end{align}

The main purpose of this note is a suitable definition of a modified field strength for Rarita-Schwinger type fields whose components are not constrained by the field equations. We denote the components of the modified field strength 
by $W_{\mu\nu}{}^\ua$ and define it in terms of the usual field strength according to
 \begin{align}
  W_{\mu\nu}=\tfrac{\Dim-3}{\Dim-1}\,T_{\mu\nu}
	-\tfrac{2(\Dim-3)}{(\Dim-1)(\Dim-2)}\,T_{\rho[\mu}\gam_{\nu]}{}^\rho
	-\tfrac{1}{(\Dim-1)(\Dim-2)}\,T_{\rho\sigma}\gam^{\rho\sigma}{}_{\mu\nu}\,,
	\label{eq3} 
 \end{align}
where $\gam^{\mu\nu}=\gam^{[\mu}\gam^{\nu]}$ and 
$\gam^{\mu\nu\rho\sigma}=\gam^{[\mu}\gam^\nu\gam^{\rho}\gam^{\sigma]}$. We shall now comment on this definition.

Rarita-Schwinger type fields are particularly relevant to supergravity theories (see 
\cite{VanNieuwenhuizen:1981ae} for a review) where they are the gauge fields of supersymmetry transformations and are termed gravitino fields. The supersymmetry transformation of a gravitino field is
 \begin{align}
  \delta_\xi \psi_\mu{}^\ua=(D_\mu \,\xi)^\ua+\ldots\, ,
	\label{eq12} 
 \end{align}
where $\xi$ is an arbitrary spinor field with components $\xi^\ua$ (`gauge parameters'), and $(D_\mu\, \xi)^\ua$ are the components of a covariant derivative of $\xi$ defined by means of a (supercovariant) spin connection $\om_\mu{}^{\nu\rho}$,
 \begin{align}
  D_\mu \xi=\6_\mu\xi-\tfrac 14\,\om_\mu{}^{\nu\rho}\,\xi\,\gam_{\nu\rho}\, .
	\label{eq13} 
 \end{align}
Accordingly, the (supercovariant) gravitino field strength $T_{\mu\nu}=D_\mu\psi_\nu{}-D_\nu\psi_\mu{}+\ldots$
transforms under supersymmetry according to
 \begin{align}
  \delta_\xi T_{\mu\nu}=\com{D_\mu}{D_\nu}\,\xi+\ldots=-\tfrac 14\, R_{\mu\nu}{}^{\rho\sigma}\,\xi\,\gam_{\rho\sigma}+\ldots\, ,
	\label{eq14} 
 \end{align}
where $R_{\mu\nu}{}^{\rho\sigma}=\6_\mu\om_\nu{}^{\rho\sigma}-\6_\nu\om_\mu{}^{\rho\sigma}+\ldots$ denote the components of the (supercovariantized) Riemann tensor. This implies that the supersymmetry transformation of the modified gravitino field strength defined 
according to \eqref{eq3} in terms of the (supercovariant) gravitino field strength is
 \begin{align}
  \delta_\xi W_{\mu\nu}=-\tfrac 14\, C_{\mu\nu}{}^{\rho\sigma}\,\xi\,\gam_{\rho\sigma}+\ldots\, ,
	\label{eq15} 
 \end{align}
where $C_{\mu\nu}{}^{\rho\sigma}$ are the components of the (supercovariantized) Weyl tensor 
 \begin{align}
  C_{\mu\nu}{}^{\rho\sigma}= 
	R_{\mu\nu}{}^{\rho\sigma}
	+\tfrac{2}{\Dim-2} \, (\delta_{[\mu}^{\rho}R^{\vphantom{\rho}}_{\nu]}{}_{\vphantom{[\mu}}^{\sigma}
	-\delta_{[\mu}^{\sigma}R^{\vphantom{\sigma}}_{\nu]}{}_{\vphantom{[\mu}}^{\rho})
	-\tfrac{2}{(\Dim-1)(\Dim-2)}\,R\, \delta_{[\mu}^{\rho}\delta_{\nu]}^{\sigma},
	\label{eq16} 
 \end{align}
where $R_{\mu}{}^{\nu}=R_{\mu\rho}{}^{\rho\nu}$ and $R=R_{\mu}{}^{\mu}$ denote the (supercovariantized) Ricci tensor
and the (supercovariantized) Riemann curvature scalar, respectively. The modified gravitino field strength is thus related by supersymmetry to the Weyl tensor but not to the Ricci tensor.
Notice also that, like the Weyl tensor, the modified field strength \eqref{eq3} vanishes for $\Dim=3$.

In order to further discuss the modified field strength \eqref{eq3} we write it as
  \begin{align}
  W_{\mu\nu}{}^\ua=T_{\rho\sigma}{}^\ub\, P^{\rho\sigma}{}_{\mu\nu\ub}{}^\ua,
	\label{eq4} 
 \end{align}
where $P^{\rho\sigma}{}_{\mu\nu\ub}{}^\ua$ are the entries of the matrix
	\begin{align}
	P^{\rho\sigma}{}_{\mu\nu}=
	\tfrac{\Dim-3}{\Dim-1}\,\delta_{[\mu}^\rho\delta_{\nu]}^\sigma\unit 
	+\tfrac{\Dim-3}{(\Dim-1)(\Dim-2)} \, 
	(\delta_{[\mu}^{\rho}\gam^{\vphantom{\rho}}_{\nu]}{}_{\vphantom{[\mu}}^{\sigma}
	-\delta_{[\mu}^{\sigma}\gam^{\vphantom{\sigma}}_{\nu]}{}_{\vphantom{[\mu}}^{\rho})
	-\tfrac{1}{(\Dim-1)(\Dim-2)}\,\gam^{\rho\sigma}{}_{\mu\nu}\,,
	\label{eq5} 
 \end{align}
where $\unit$ is the ${2^{\lfloor \Dim/2 \rfloor}}\times {2^{\lfloor \Dim/2 \rfloor}}$ unit matrix. 
This implies
\begin{align}
	P_{\mu\nu}{}^{\rho\sigma}=
	\tfrac{\Dim-3}{\Dim-1}\,\delta_{[\mu}^\rho\delta_{\nu]}^\sigma\unit 
	-\tfrac{\Dim-3}{(\Dim-1)(\Dim-2)} \, 
	(\delta_{[\mu}^{\rho}\gam^{\vphantom{\rho}}_{\nu]}{}_{\vphantom{[\mu}}^{\sigma}
	-\delta_{[\mu}^{\sigma}\gam^{\vphantom{\sigma}}_{\nu]}{}_{\vphantom{[\mu}}^{\rho})
	-\tfrac{1}{(\Dim-1)(\Dim-2)}\,\gam_{\mu\nu}{}^{\rho\sigma}.
	\label{eq5a} 
 \end{align}
These matrices fulfill the identities
\begin{align}
   &P^{\rho\sigma}{}_{\mu\nu}\gam^{\mu\nu\kappa}=0, \quad
	 \gam^{\mu\nu\kappa} P_{\mu\nu}{}^{\rho\sigma}=0,\label{eq6} \\
   &P^{\rho\sigma}{}_{\kappa\lambda} P^{\kappa\lambda}{}_{\mu\nu}=P^{\rho\sigma}{}_{\mu\nu}\,, \label{eq7} \\
	 &P^{\rho\sigma}{}_{\mu\nu}{}^\top=\CC P_{\mu\nu}{}^{\rho\sigma}\IC,
	\label{eq8} 
 \end{align}
where in \eqref{eq8} $P^{\rho\sigma}{}_{\mu\nu}{}^\top$ denotes the transpose of $P^{\rho\sigma}{}_{\mu\nu}$, 
and $\CC$ and $\IC$ denote a charge conjugation matrix and its inverse which relate the gamma matrices to the transpose gamma matrices 
according to $\gam^\mu{}^\top=-\eta\CC \gam^\mu\IC$ with $\eta\in\{+1,-1\}$
and are used to raise and lower spinor indices according to
\begin{align}
  W_{\mu\nu\ua}=\IC{}_{\ua\ub}\,W_{\mu\nu}{}^\ub,\quad W_{\mu\nu}{}^\ua=\CC^{\ua\ub}\,W_{\mu\nu\ub}\,.
	\label{eq9} 
 \end{align}
The first equation \eqref{eq6} implies that the modified field strength {\em identically} fulfills 
$W_{\mu\nu}\gam^{\mu\nu\rho}=0$, i.e. indeed the components of the modified field strength are not constrained by the field equations. 
\eqref{eq7} implies that the matrices
$P^{\rho\sigma}{}_{\mu\nu}$ define a projection operation on the field strength
as they define an idempotent operation. Hence, this operation projects to linear combinations of components of the usual field strength which are not constrained by the field equations.
Owing to \eqref{eq8} the components of the modified field strength with lowered spinor index are related to the usual field strength according to
\begin{align}
   W_{\mu\nu\ua}=P_{\mu\nu}{}^{\rho\sigma}{}_\ua{}^\ub\,T_{\rho\sigma\ub}\,.
	\label{eq10} 
 \end{align}
The second equation \eqref{eq6} thus implies that the modified field strength with lowered spinor index identically fulfills $\gam^{\mu\nu\rho}W_{\mu\nu}=0$.

Moreover, the modified field strength coincides with the usual field strength on-shell whenever $T_{\mu\nu}\gam^{\mu\nu\rho}$ vanishes on-shell:
\begin{align}
   T_{\mu\nu}\gam^{\mu\nu\rho}\approx 0 \quad \Then\quad
	  W_{\mu\nu}\approx T_{\mu\nu}\,.
	\label{eq19} 
 \end{align}
This is obtained from \eqref{eq3} by means of the following implications of 
$T_{\mu\nu}\gam^{\mu\nu\rho}\approx 0$:\footnote{See also, e.g., equations (35) of \cite{Cremmer:1980ru}.}
\begin{align}
   &T_{\mu\nu}\gam^{\mu\nu\rho}\approx 0 \ \STACK{\cdot\gam_\rho}{\Then}\
	  T_{\mu\nu}\gam^{\mu\nu}\approx 0\ \STACK{\cdot\gam_\rho}{\Then}\
		T_{\rho\mu}\gam^{\mu}\approx 0\ \STACK{\cdot\gam_\sigma}{\Then}\
		T_{\mu[\rho}\gam_{\sigma]}{}^{\mu}\approx -T_{\rho\sigma}\, ,
	 \label{eq17} \\
   &T_{\mu\nu}\gam^{\mu\nu}\approx 0\, \STACK{\cdot\gam_{\rho\sigma}}{\Then}\,
	  T_{\mu\nu}\gam^{\mu\nu}{}_{\rho\sigma}-4T_{\mu[\rho}\gam_{\sigma]}{}^{\mu}-2T_{\rho\sigma}\approx 0
		\, \STACK{\eqref{eq17}}{\Then}\, 
		T_{\mu\nu}\gam^{\mu\nu}{}_{\rho\sigma}\approx -2T_{\rho\sigma}\, .
	\label{eq18} 
 \end{align}
In particular, \eqref{eq19} applies to free Rarita-Schwinger type gauge fields satisfying 
$T_{\mu\nu}\gam^{\mu\nu\rho}\approx 0$ with $T_{\mu\nu}=2\6_{[\mu}\psi_{\nu]}$. Hence, \eqref{eq19} also applies to linearized supergravity theories. Moreover \eqref{eq19} applies
to several supergravity theories at the full (nonlinear) level, such as to $N=1$ pure supergravity in $\Dim=4$ 
\cite{Freedman:1976xh,Deser:1976eh} and to supergravity 
in $\Dim=11$ \cite{Cremmer:1978km}, where the equations of motion imply $T_{\mu\nu}\gam^{\mu\nu\rho}\approx 0$ for the 
usual supercovariant gravitino field strengths $T_{\mu\nu}$. 
Accordingly, in these cases the modified gravitino field strength fulfills on-shell the same
Bianchi identities as the usual gravitino field strength, 
such as ${\cal D}_{[\mu}W_{\nu\rho]}\approx 0$ in $N=1$, $D=4$ pure supergravity for the supercovariant derivatives of $W_{\mu\nu}$.

If $T_{\mu\nu}\gam^{\mu\nu\rho}\approx X^\rho$ 
for some $X^\rho$, one obtains in place of equations \eqref{eq17} and \eqref{eq18} the following implications:
\begin{align}
   T_{\mu\nu}\gam^{\mu\nu\rho}\approx  X^\rho \ \Then\
		&T_{\mu[\rho}\gam_{\sigma]}{}^{\mu}\approx -T_{\rho\sigma}-\tfrac{1}{2(\Dim-2)}X^\mu\gam_{\mu\rho\sigma}
		+\tfrac{\Dim-4}{2(\Dim-2)}X_{[\rho}\gam_{\sigma]}\, ,
	 \label{eq17a} \\
   &	T_{\mu\nu}\gam^{\mu\nu}{}_{\rho\sigma}\approx -2T_{\rho\sigma}
	 -\tfrac{1}{\Dim-2}X^\mu\gam_{\mu\rho\sigma}
	 +\tfrac{2(\Dim-3)}{\Dim-2}X_{[\rho}\gam_{\sigma]}\, .
	\label{eq18a} 
 \end{align} 
Using \eqref{eq17a} and \eqref{eq18a} in \eqref{eq3} gives in place of \eqref{eq19}:
 \begin{align}
   &T_{\mu\nu}\gam^{\mu\nu\rho}\approx X^\rho \ \Then\
	W_{\mu\nu}\approx T_{\mu\nu}+X_{\mu\nu}\,,\label{eq19a} \\
	&X_{\mu\nu}=\tfrac{1}{(\Dim-1)(\Dim-2)}X^\rho\gam_{\rho\mu\nu}
	-\tfrac{\Dim-3}{(\Dim-1)(\Dim-2)}X_{[\mu}\gam_{\nu]}
	\label{eq19b} 
 \end{align} 
which implies ${\cal D}_{[\mu}W_{\nu\rho]}\approx {\cal D}_{[\mu}T_{\nu\rho]}+{\cal D}_{[\mu}X_{\nu\rho]}$,
relating the on-shell Bianchi identities for the modified gravitino field strength 
to those for the usual gravitino field strength. I remark that $X_{\mu\nu}$ fulfills 
$X_{\mu\nu}\gam^{\mu\nu\rho}=-X^\rho$ (identically). Hence, $\hat W_{\mu\nu}=W_{\mu\nu}-X_{\mu\nu}$ fulfills
$\hat W_{\mu\nu}\approx T_{\mu\nu}$ and $\hat W_{\mu\nu}\gam^{\mu\nu\rho}=X^\rho$, and may be used in place of  
$W_{\mu\nu}$ as a modified field strength that coincides on-shell with the usual field strength in the case
$T_{\mu\nu}\gam^{\mu\nu\rho}\approx X^\rho$.
Moreover, $\hat T_{\mu\nu}=T_{\mu\nu}+X_{\mu\nu}$ fulfills $\hat T_{\mu\nu}\gam^{\mu\nu\rho}\approx 0$ if
$T_{\mu\nu}\gam^{\mu\nu\rho}\approx X^\rho$, i.e., alternatively one may redefine $T_{\mu\nu}$ 
to $\hat T_{\mu\nu}$ and use $W_{\mu\nu}=\hat T_{\rho\sigma}P^{\rho\sigma}{}_{\mu\nu}$ which then coincides 
on-shell with the redefined field strength, i.e.,
$W_{\mu\nu}\approx \hat T_{\mu\nu}$.

\eqref{eq19} and the identities 
$W_{\mu\nu}\gam^{\mu\nu\rho}=0$ also show that the matrices
$P^{\rho\sigma}{}_{\mu\nu}$ remove precisely those linear combinations of components from 
the usual field strength which occur in $T_{\mu\nu}\gam^{\mu\nu\rho}$. For instance, in $N=1$ pure supergravity in $\Dim=4$ the modified gravitino field strength $W_{\mu\nu}$ 
contains precisely those linear combinations of components of 
$T_{\mu\nu}$ which, using van der Waerden notation with $\ua=1,2,\dot 1,\dot 2$, are denoted by 
$W_{\alpha\beta\gamma}|$ and $\overline{W}_{\dot\alpha\dot\beta\dot\gamma}|$ in the chapters XV and XVII of \cite{Wess:1992cp}
but no linear combination occurring in $T_{\mu\nu}\gam^{\mu\nu\rho}$.
The modified gravitino field strengths are thus particularly useful for the construction and classification of on-shell invariants, counterterms and consistent deformations of supergravity theories.

We end this note with the remark that there is an analog of the modified field strength \eqref{eq3} for first order derivatives of Dirac type spinor fields $\psi$ which are subject to field equations which read or imply 
$T_\mu\gam^\mu+\ldots\approx 0$ with $T_\mu=\6_\mu\psi+\ldots$. This analog is defined according to
\begin{align}
   W_{\mu}=\tfrac{\Dim-1}{\Dim}\,T_\mu-\tfrac{1}{\Dim}\,T_\nu\gam^\nu{}_\mu =T_\nu P^\nu{}_\mu\,,\quad
	P^\nu{}_\mu=\tfrac{\Dim-1}{\Dim}\,\delta_\mu^\nu\unit-\tfrac{1}{\Dim}\,\gam^\nu{}_\mu\,.
	\label{eq20} 
 \end{align}
The matrices $P^\nu{}_\mu$ fulfill identities analogous to \eqref{eq6}-\eqref{eq8}:
\begin{align}
   P^\nu{}_\mu \gam^\mu=0, \quad
	 \gam^{\mu} P_{\mu}{}^{\nu}=0,\quad
   P^\nu{}_\rho P^\rho{}_\mu=P^\nu{}_\mu\,,\quad
	 P^\nu{}_\mu{}^\top=\CC P_{\mu}{}^{\nu}\IC.
	\label{eq21} 
 \end{align}
In particular $W_\mu$ thus identically fulfills $W_{\mu}\gam^\mu=0$.
Furthermore, analogously to \eqref{eq19}, $T_{\mu}\gam^{\mu}\approx 0$ implies $W_{\mu}\approx T_{\mu}$
because $T_{\mu}\gam^{\mu}\approx 0$ implies $T_\nu\gam^\nu{}_\mu\approx -T_\mu$. Hence, the matrices $P^\nu{}_\mu$ remove precisely those linear combinations of components from $T_\mu$ which occur in $T_\mu\gam^\mu$. For $W_\mu$ with lowered spinor index one has $W_\mu=P_{\mu}{}^{\nu}T_\nu$ and $\gam^\mu W_\mu=0$.

{\bf Added note:} After this paper was published the author learned that $\Dim=5$ versions of the modified gravitino field strength \eqref{eq3} and of the projection operator \eqref{eq5} were published already in \cite{Butter:2014xxa}.

\end{document}